\documentstyle[12pt]{article}
\begin{document}

\title{Evidence for vortex staircases in the whole angular range due to
competing correlated pinning mechanisms}
\author{A. Silhanek$^1$, L. Civale$^1$, S. Candia$^1$, G. Nieva$^1$, G.
Pasquini$^2$ and H. Lanza$^2$.\\
$^1$Comisi\'{o}n Nacional de Energ\'{\i}a At\'{o}mica-Centro At\'{o}mico\\
Bariloche and Instituto Balseiro, 8400 Bariloche, Argentina. \\
$^2$Comisi\'{o}n Nacional de Energ\'{\i}a At\'{o}mica-Departamento de\\
F\'{\i}sica, Av. del Libertador 8250, 1429, Buenos Aires, Argentina. }
\date{November 30, 1998}
\maketitle

\begin{abstract}
We analyze the angular dependence of the irreversible
magnetization of YBa$_2$Cu$_3$O$_7$ crystals with columnar defects inclined
from the c-axis. At high fields a sharp maximum centered at
the tracks' direction is observed. At low fields we identify a lock-in phase
characterized by an angle-independent pinning strength and observe
an angular shift of the peak towards the c-axis that originates in the
material anisotropy. The interplay among columnar defects, twins and ab-planes 
generates a variety of staircase structures. We show that correlated pinning
dominates for all field orientations.

\end{abstract}



A difficult aspect of the study of vortex dynamics in HTSC in the presence
of correlated disorder is the determination of flux structures for applied
fields tilted with respect to the pinning potential. As 3D vortex
configurations cannot be directly observed, our knowledge is mostly based
on the analysis of the angular dependence of magnetization, susceptibility
or transport data\cite{civale-91a,kwok-92,grigorieva-92,fleshler-93,li-93,klein-93,hardy-96,oussena-96,zhukov-97b,herbsommer-98}.

According to theoretical models\cite{nelson-92,blatter-94}, when the angle
between the applied field ${\bf H}$ and the defects is smaller than the
lock-in angle $\varphi _L$ vortices remain locked into the defects thus
producing a transverse Meissner effect. For tilt angles larger than $\varphi
_L$ and smaller than a trapping angle $\varphi _T$, vortices form staircases
with segments pinned into different defects and connected by unpinned or
weakly pinned kinks. Beyond $\varphi _T$, vortices will be straight and take
the direction of the applied field, thus being unaffected by the correlated
nature of the pinning. In principle, this picture should apply with minor
differences to twins, columnar defects and intrinsic pinning\cite{blatter-94}.

Many experiments have confirmed the directional pinning due to columnar
defects, twins and Cu-O planes\cite{civale-91a,kwok-92,grigorieva-92,fleshler-93,li-93,klein-93,hardy-96,oussena-96,zhukov-97b,herbsommer-98}. Evidence for 
a locked-in phase arises
from the observation of the transverse Meissner effect\cite{zhukov-97b}, but
a quantitative determination of $\varphi _L(H,T)$ for columnar defects had
not been done until now. The introduction of columnar defects inclined with
respect to the c-axis has been used\cite{civale-91a,klein-93,hardy-96,zhukov-97b,herbsommer-98} to discriminate
their pinning effects from those due to twin boundaries, and from anisotropy
effects. However, the combined effect of the various correlated structures on the vortex configurations remain largely unexplored.

In this work we report studies of the vortex pinning in YBa$_2$Cu$_3$O$_7$ crystals with
inclined columnar defects, for the whole range of field orientations. This
allows us to determine the misalignment between the applied and internal
fields due to anisotropy, as well as to identify the angular range of
influence of each correlated pinning structure. We present the first
determination of the lock-in angle of tracks using irreversible
magnetization.

The crystal used in this study was grown by the self flux method\cite{delacruz-94},
and has dimensions $\sim 200\times 600\times 8.5\mu m^3$. Columnar defects
at an angle $\Theta _D\approx 32^{\circ }$ from the c-axis and a density
corresponding to a matching field $B_\Phi=3T$ were introduced by
irradiation with $315$ $MeV$ $Au^{23+}$ ions at the Tandar accelerator 
(Buenos Aires, Argentina).

DC magnetization ${\bf M}$ was measured in a Quantum Design SQUID
magnetometer with two sets of pick up coils, and both the longitudinal ($M_l$%
, parallel to ${\bf H}$) and transverse ($M_t$, perpendicular to ${\bf H}$)
components were recorded. The sample could be rotated {\it in situ} around an axis
perpendicular to ${\bf H}$ using a home-made device. The angle $%
\Theta$ between the normal to the crystal (that coincides with the
c-axis) and ${\bf H}$ was determined with absolute accuracy $\sim 1^{\circ }$, and relative variations between
adjacent angles better than $0.2^{\circ }$. The details of the procedure are 
described elsewhere\cite{casa}.

Magnetization loops $M_l(H)$ and $M_t(H)$ were recorded at fixed $T$ and $\Theta$. 
Sample was then rotated, warmed up above $T_c$ and cooled
down in zero field to start a new run. We use the hysteresis widths $%
\Delta M_l\left( H\right) $ and $\Delta M_t\left( H\right) $ to calculate
the modulus $M_{i}=\frac 12\sqrt{\Delta M_l^2+\Delta M_t^2}$ and direction
of the irreversible magnetization vector ${\bf M}_{i}$. It is known that
in thin samples ${\bf M}_{i}$ is normal to the surface due to geometrical
constrains\cite{casa,zhukov-97a}, except above a critical angle (of $\sim 87^{\circ }$ 
for our crystal). We confirmed that 
${\bf M}_{i}\parallel c$ within $1^{\circ }$, for all $\Theta<85^{\circ }$.

From now on we analyze the modulus $M_{i}$ as a function of $T$, $H$ and $%
\Theta$. Figure 1 shows $M_{i}(\Theta)$ at $60$ and $70K$ and several 
values of $H$. According to the Bean model, $M_{i} \propto J$, where the 
screening current density $J$ is lower than $J_c$ due to thermal
relaxation. The geometrical factor $M_{i}/J$ depends on $%
\Theta$, but is almost constant for $\Theta$ below the critical angle.

The most obvious feature of Fig. 1 is the asymmetry with respect to the
c-axis, which is due to the uniaxial pinning of the inclined
tracks. At high fields ($H\geq 1T$) we observe a large peak in the
direction of the tracks $\Theta _D\approx 32^{\circ }$. For $H<1T$ the
peak becomes broader and {\it progressively shifts away from the tracks in
the direction of the c-axis} as $H$ decreases. The shift decreases with increasing $T$
as shown in figure 2, where the angle $\Theta_{\max }$ of the maximum in $%
M_{i}$ is plotted as a function of $H$ for three temperatures. The inset
of figure 2 shows a blow-up of the data of fig. 1 for $H=0.4T$ and $T=60K$. This curve
exhibits the second main characteristic of the low field results, namely the
existence of a {\it plateau} in $M_{i}\left( \Theta\right) $ (We define $\Theta_{\max }$ as the center of the plateau).

We first discuss the origin of the shift. Maximum pinning is expected to
occur when the tracks are aligned with the direction that {\it the vortices
would have in the absence of pinning.} For an anisotropic material, such
direction {\it does not coincide} with ${\bf H}$. If $\Theta _B$ is the
angle between the equilibrium induction field ${\bf B}$ (which represents
the vortex direction) and the c-axis, minimization of the free energy for
$H_{c1}^c\ll H\ll H_{c2}^c$ gives\cite{blatter-94}

\begin{equation}
\sin \left( \Theta _B-\Theta\right) \approx \frac{H_{c1}^c(1-\varepsilon ^2)}
{2H\ln \kappa }\frac{\sin \Theta _B\cos \Theta _B}{\varepsilon \left( \Theta _B\right) }%
\ln \left( \frac{H_{c2}(\Theta _B)}{B}\right)  \label{desviacion}
\end{equation}
where $H_{c2}(\Theta_B)= {H_{c2}^c}/{\varepsilon(\Theta _B)}$. Here $H_{c1}^c$ and $H_{c2}^c$
are the lower and upper c-axis critical fields, $\varepsilon $ is the
anisotropy and $\varepsilon \left( \theta \right) =\left( \cos ^2\theta +\varepsilon ^2\sin
^2\theta \right) ^{1/2}$. For $\varepsilon <1$ vortices tilt towards the {\it ab} plane. When
$\Theta=\Theta _D$ we have $\Theta _B>\Theta _D$ and the optimum pinning situation is not
satisfied. Instead, maximum $M_{i}$ occurs at the vortex-track alignment condition $\Theta
_B=\Theta _D$. This corresponds to an applied field angle $\Theta_{\max }<\Theta _D$ that can
be calculated from Eq. \ref{desviacion} by setting $\Theta _B=\Theta _D\approx 32^{\circ }$.
(In this picture the peak cannot occur at $ \Theta<0$, thus the negative values of
$\Theta_{\max }$ given by Eq. \ref{desviacion} at low $H$ are unphysical, and $\Theta_{\max
}\rightarrow 0$ as $H\rightarrow 0$).

The solid lines in fig. 2 are fits to Eq. \ref{desviacion} with {\it %
fixed} parameters\cite{blatter-94} $\varepsilon=1/7$ and $H_{c2}^c(T)=1.6T/K\times
(T_c-T)$ (the fits are not very sensitive to any of them). Using $%
H_{c1}^c(T)/2\ln \kappa =\Phi _0/8\pi \lambda _{ab}^2(T)$ and $\lambda
_{ab}^2(T)\approx \lambda _{ab}^2(0)(1-T/T_c)^{-1}$, we obtain a good fit to
the data as a function of field and temperature by setting only one free
parameter, $\lambda _{ab}(0)\approx 500\AA $. Although this value is
significantly smaller than the accepted value\cite{blatter-94} ($\sim 1400\AA $), we
nevertheless consider that this simple model captures the basic physics. 
We note that Zhukov et al. \cite{zhukov-97b} have
reported lock-in angles for twin boundaries in YBa$_2$Cu$_3$O$_7$ crystals that imply an 
$H_{c1}^c$ about 5 times larger than the usual values, a result suggestively
similar to our case.

We now return to the plateau seen in the inset of figure 2. The constancy of 
$M_{i}\left( \Theta\right) $ indicates that the pinning energy remains
constant and equal to the value at the alignment condition $\Theta
_B=\Theta _D$. This behavior is a fingerprint of the lock-in phase\cite{nelson-92}. 
The extension of the plateau in the $H-\Theta$
plane at $60K$ (determined with accuracy $\sim 1^{\circ }$) is shown as bars in Fig. 2. Its width decreases approximately
as $H^{-1}$, as expected\cite{nelson-92} for $\varphi _L$, and for $H>1T$ it becomes
undetectable with our resolution. The decrease of $M_{i}$ at the edges of
the plateau is sharp, a result consistent with the appearance of kinks,
which not only reduce $J_c$ but also produce a faster
relaxation.

When $\left| \Theta _B-\Theta _D\right| >\varphi _L$ vortices form
staircases. Two questions arise here. First, which is the direction of the
kinks that connect the pinned portions of the vortices? Second, do we
observe evidence for a trapping angle $\varphi _T$?

For $\Theta>\Theta_{\max }$, there is a wide angular range in Fig. 1
in which $M_{i}\left( +\Theta\right) >M_{i}\left( -\Theta\right) $
for all $H$, i.e., pinning is stronger when $H$ is closer to the tracks than
in the crystallographically equivalent configuration in the opposite side.
This asymmetry demonstrates that at the angle $+\Theta$ vortices form
staircases, with segments trapped in the tracks. For $\Theta<\Theta_{\max }$
we again observe asymmetry, $M_{i}\left( \Theta\right) $
crosses $\Theta=0$ with positive slope, indicating that pinning decreases
as $H$ is tilted away from the tracks. We can conclude that staircases
extend at least beyond the c-axis into the $\Theta<0$ region.

The angle $\theta _k$ between the kinks and the c-axis can be calculated 
by minimization of the free energy\cite{blatter-94}. For simplicity, we 
consider the case $H\gg H_{c1}^c$, where $\Theta _B=\Theta$
and the problem reduces to calculate the energy of one vortex, as the
other terms in the free energy are the same for all configurations\cite{hardy-96,herbsommer-98}.
If $L_p$ is the length of a pinned segment, and $L_k$ the length of the kink
(see sketch in figure 4), the energy is $E\propto L_p\epsilon _p\left(
\Theta _D\right) +L_k\epsilon _f\left( \theta _k\right) $, where $\epsilon
_f\left( \theta _k\right) \approx \varepsilon _0\varepsilon \left( \theta
_k\right) \left[ \ln \kappa +0.5\right] $ and $\epsilon _p\left( \Theta
_D\right) \approx \varepsilon _0\varepsilon \left( \Theta _D\right) \left[
\ln \kappa +\alpha _t\right] $ are the line energy for free and pinned
vortices respectively, $\varepsilon _0$ is the vortex energy scale and $%
\alpha _t<0.5$ parametrizes the core pinning energy due to the tracks
(smaller $\alpha _t$ implies stronger pinning). Minimizing $E$ with respect
to $\theta _k$ we obtain two orientations, $\theta _k^{-}$ for $%
\Theta<\Theta _D$ and $\theta _k^{+}$ for $\Theta>\Theta _D$.

As the tracks are inclined, $\left|\theta _k^{-}\right|$ and 
$\left|\theta _k^{+}\right|$ are
different. However, those angles are independent of $\Theta$. As $\left|
\Theta-\Theta _D\right| $ increases, $\theta _k^{\pm }$ remain constant
while $L_p$ decreases and the number of kinks increases, consequently the
pinning energy lowers. This accounts for an $M_{i}$ that decreases as we
move away from the tracks. In particular, for $\Theta=\theta _k^{\pm }$
vortices become straight ($L_p=0$), thus $\varphi _T^{\pm }=\left| \theta
_k^{\pm }-\Theta _D\right| $ are the trapping angles in both directions. In
general $\theta _k^{\pm }$ must be obtained numerically, but for $%
\varepsilon \tan \theta _k\ll 1$ and $\varepsilon \tan \Theta _D\ll 1$ we
obtain 
\begin{equation}
\tan \theta _k^{\pm }\approx \tan \Theta _D\pm \frac 1\varepsilon \sqrt{%
\frac{1-2\alpha _t}{\ln \kappa +0.5}}  \label{kink}
\end{equation}

Eq. \ref{kink} adequately describes the main features of the asymmetric
region in Fig. 1, and for $\Theta _D=0$ it coincides with the usual
estimates\cite{nelson-92,blatter-94} of $\varphi _T$.

There is, however, an important missing ingredient in the standard
description presented above, namely the existence of twins and Cu-O layers,
which are additional sources of correlated pinning. This raises the
possibility that vortices may simultaneously adjust to more than one of
them, forming different types of staircases.

Pinning by twin boundaries is visible in figure 1 as an additional peak centered at the c-axis
for $H=2T$ and $T=60K$. A blow-up of that peak is shown in the inset. We observe this
maximum for $H\geq 1T$. The width of this peak, $\sim 5^{\circ }$, is in the typical range
of reported trapping angles for twins \cite
{grigorieva-92,fleshler-93,li-93,oussena-96,zhukov-97b}. On the other hand, the fact that the
peak is mounted over an inclined background implies that vortices are also trapped by the
tracks. Thus, vortices in this angular range contain segments both in the tracks
and in the twins. These two types of segments are enough to build up the staircases for $\Theta>0$, but for $\Theta<0$ a third group of inclined kinks with $\theta _k<0$ must exist in order to have vortices parallel to ${\bf H}$.

Another fact to be considered is that there is an angle $\Theta_{sym}$ (which is only weakly dependent on $H$) beyond which 
$M_{i}\left( \Theta\right)$ recovers the symmetry with respect to the c-axis. This is illustrated in Fig. 3, where $M_{i}$ data for $-\left| \Theta\right| $ was reflected along the c-axis and
superimposed to the results for $+\left| \Theta\right| $. 

One possible interpretation is that for $\Theta>\Theta_{sym}$
staircases disappear, i.e., that $\Theta_{sym}=\theta _k^{+}$ and we are
determining $\varphi _T^{+}=\Theta_{sym}-\Theta _D$. However, this is
inconsistent with our experimental results. Indeed, $\varphi _T^{+}$ should
decrease with $T$, and this decrease should be particularly strong above the
depinning temperature \cite{krusin-elbaum-96} $T_{dp}\sim 40K$ due to the
reduction of the pinning energy by entropic effects \cite{nelson-92}%
. This expectation is in sharp contrast with the observed increase of $%
\Theta_{sym}$ with $T$, which is shown in Figure 4 for $H=2T$%
. Thus, the interpretation of $\Theta_{sym}$ as a measure of the trapping
angle is ruled out. Moreover, if in a certain angular range vortices were
not forming staircases, pinning could be described by a scalar disorder,
then at high fields $M_{i}\left( \Theta\right) $ should
follow the anisotropy scaling law\cite{blatter-92} $M_{i}\left( H,\Theta
\right) =M_{i}\left( \varepsilon \left( \Theta\right) H\right) $.
Consistently, we do not observe such scaling in any angular range.

Our alternative interpretation is that, at large ${\Theta}$, the kinks become 
trapped by the {\it ab}-planes. This idea has
been used by Hardy et al. \cite{hardy-96} to explain that the $J_c$ at low $T
$ in the very anisotropic Bi and Tl compounds with tracks at $\Theta
_D=45^{\circ }$ was the same for either $\Theta=45^{\circ }$ or $\Theta=-45^{\circ }$. 
Our situation is different, as we are comparing two kinked configurations.

We first note that, according to Eq. \ref{kink}, $\theta _k^{\pm }$ cannot be exactly $90^{\circ }$ for finite $\varepsilon $, thus the intrinsic pinning must be incorporated into the model by assigning a lower energy to kinks in the {\it ab}-planes. Vortices may now form structures consisting of segments trapped in the columns connected by segments trapped in the {\it ab}-planes, or alternatively an inclined kink may transform into a staircase of smaller kinks connecting segments in the planes (see sketches in figure 4). We should now compare the energy of the new configurations with that containing kinks at angles $\theta _k^{\pm }$. This is equivalent to figure out whether the kinks at $\theta _k^{\pm }$ lay within the trapping regime for the planes or not. The problem with this analysis is that, as $\theta _k^{\pm }$ are independent of $\Theta$, one of the two possibilities (either inclined or trapped kinks), will be the most favorable for all $\Theta$. Thus, this picture alone cannot explain the crossover from an asymmetric to a symmetric regime in $M_{i}\left( \Theta\right) $.

The key additional concept in this scenario is the dispersion in the pinning energy. The angles $\theta _k^{\pm }$ depend on the pinning strength of the adjacent tracks ($\alpha _t$ in Eq. \ref{kink}), thus dispersion in $\alpha _t$ implies dispersion in $\theta _k^{\pm }$. As $\Theta$ increases, it becomes larger than the smaller $\theta _k^{\pm }$'s (that connect the weaker defects) and the corresponding kinks disappear. The vortices involved, however, do not become straight, but remain trapped by stronger pins connected by longer kinks with larger $\theta _k^{\pm }$. This process goes on as $\Theta$ grows: the weaker tracks progressively become unnefective as the ''local'' $\theta _k$ is exceeded, and the distribution of $\theta _k^{\pm }$ shifts towards the {\it ab}-planes. When a particular kink falls within the trapping angle of the planes, a switch to the pinned-kink structure occurs. In this picture, the gradual crossover to the symmetric regime takes place when most of the remaining kinks are pinned by the planes.

If kinks become locked, the total length
of a vortex that is trapped inside columnar defects is the total length of a
track, independent of $\Theta$, and the total length of the kinks is $%
\propto \tan \left( \left| \Theta\pm \Theta _D\right| \right) $ for field
orientations $\pm \Theta$ respectively. As $\left| \Theta\right| $
grows, the relative difference between the line energy in both orientations
decreases, an effect that is reinforced by the small line energy of the
kinks in the {\it ab}-planes. If kinks are not locked but rather form
staircases, taking into account that the trapping angle for the ab-planes is
small\cite{fleshler-93} ($\sim 5^{\circ }$), the same argument still applies to a good
approximation. The temperature dependence of $\Theta_{sym}$ is now easily
explained by a faster decrease of the pinning of the {\it ab}-planes with $T$
as compared to the columnar defects. Additional evidence in support of our description 
comes from recent transport measurements in {\it twinned} YBa$_2$Cu$_3$O$_7$ crystals, 
which show that in the liquid phase vortices remain correlated 
along the c-axis {\it for all field orientations}\cite{morre-97}, suggesting that 
they are composed solely of segments in the twins and in the 
{\it ab}-planes.

In summary, we have shown that the combined effect of the three sources of
correlated pinning must be taken into account to describe the vortex
structure in samples with inclined columnar defects. We demonstrate that the
lock-in phase exhibits an angle independent pinning strength, and show the
decrease of the lock-in angle with field. Our results show that a variety of
complex staircases are formed depending on the field orientation and
strongly suggest that, at high temperatures, correlated structures dominate
vortex pinning over random disorder in the whole angular range.

Work partially supported by ANPCyT, Argentina, PICT 97 No.01120. A.S. and G.N. are members of CONICET. We acknowledge useful discussions with S. Grigera, F. de la Cruz, E. Osquiguil and D. Niebieskikwiat.\\

\noindent
Figure 1:Irreversible magnetization $M_i$ as a function of the applied field angle $\Theta$ for several fields $H$, at temperatures (a) $70K$ and (b) $60K$. Inset: blow up of the $H=2T$ data near the c-axis for $T=60K$ (the units are the same of those in the main figure.\\

\noindent
Figure 2:Angle $\Theta_{max}$ of the maximum in $M_i(\Theta)$ as a function of H for three temperatures. The solid lines are fits to Eq.(1) (see text). Bars mark the width of the plateau. Inset: $M_i(\Theta)$ in the region of the plateau.\\

\noindent
Figure 3:Irreversible magnetization $M_i$ versus field angle $\Theta$ for three fields (curves are vertically displaced for clarity). Open symbols: data for $\Theta>0$. Solid symbols: data for $\Theta<0$, reflected with respect to the c-axis. Arrows indicate the angle $\Theta_{sym}$ beyond which the behavior is symmetric. The procedure of reflection is sketched in the inset.\\

\noindent
Figure 4:Temperature dependence of $\Theta_{sym}$ (see fig.3). The solid line is a guide to the eye. The sketches show possible vortex staircases for $\Theta > \Theta_{D}$.\\

\bibliographystyle{prsty}

\end{document}